\documentstyle[12pt,aaspp4]{article}
\begin{document}
\title{Quasar Parallax: a Method for Determining Direct
Geometrical Distances to Quasars}
\author{Martin Elvis and Margarita Karovska}
\affil{Harvard-Smithsonian Center for Astrophysics, Cambridge MA
02138, USA}
\affil{\tt revised: 5~pm, 3 November 2002}
\authoremail{elvis@cfa.harvard.edu}
%

%
\begin{abstract}
We describe a novel method to determine direct geometrical
distances to quasars that can measure the cosmological constant,
$\Lambda$, with minimal assumptions.  This method is equivalent
to geometric parallax, with the `standard length' being the size
of the quasar broad emission line region (BELR) as determined
from the light travel time measurements of reverberation mapping.
The effect of non-zero $\Lambda$ on angular diameter is large,
40\% at z=2, so mapping angular diameter distances vs. redshift
will give $\Lambda$ with (relative) ease.
In principle these measurements could be made in the UV, optical,
near infrared or even X-ray bands.  Interferometers with a
resolution of 0.01~mas are needed to measure the size of the BELR
in z=2 quasars, which appear plausible given reasonable short
term extrapolations of current technology.
\end{abstract}

\bigskip 

We propose a method to determine direct geometrical distances to
radio-quiet quasars that can measure the cosmological constant,
$\Lambda$, with minimal assumptions. The largest scale use of
astronomical distances is the measurement of the geometry of the
universe.  Distance measurements using the brightness of
supernovae of type 1a (SN1a) up to a redshift, z$\sim$1.5 as
`standard candles' suggest that the universe has a non-zero
cosmological constant, $\Lambda$, so that the expansion of the
universe is accelerating (Perlmutter et al. 1999, Riess et
al. 1998, 2001). Most of the energy density of the universe,
$\Omega$, would then lie in this repulsive `dark energy',
$\Omega_{\Lambda}\sim$0.7, rather than in `normal dark matter',
$\Omega_{m}\sim$0.3. This measurement may be key to new physics
and new cosmology (Carroll 2001). However, like most methods for
finding astronomical distances, the SN1a method has inherent
uncertainties due to model dependent assumptions, the small size
of the effect ($\sim$10\%), and difficulties of measurement
(Rowan-Robinson 2002; Wright 2002; Branch et al. 2001; Richardson
et al. 2002; Linder 2001; Goobar et al. 2002). The need for a
second, independent, means of determining $\Lambda$ is clear and
urgent.

The most direct and model independent method for finding
astronomical distances involves simple geometry. This method,
parallax (Bessel \& Rath 1839), uses the motion of the Earth
around the Sun to measure the angular displacement of a nearby
star, against an effectively unmoving distant background of
stars. By this means we use a known length, in this case the
Earth's orbital radius (1~AU), and a known angle to solve the
isosceles triangle created by these two quantities and so solve
for the distance to the star (figure~1a).
Here we propose an equivalent geometric method to determine
distances to quasars (figure~1b). In this case the triangle is
inverted, with the known length being at the distant quasar,
instead of at the Earth. The `standard length' we propose to
employ is the size of the quasar broad emission line region (BELR,
Peterson 1997), which is known from light travel time
measurements.
The angle in the triangle ($\theta$, figure~1b) is that subtended
by the same BELR, as measured with an interferometer. By
measuring both quantities to a series of quasars at different
redshifts the `angular diameter distance', $D_A$, can be mapped
out against redshift, $z$.  The resulting
angular~diameter~-~redshift relation is a basic characteristic of
the space-time metric of the universe. From this relation we can
determine $\Lambda$. An equivalent method was used to measure the
distance to SN1987a (Panagia et al. 1991, Binney and Merrifield
1998) to 6\%.
Because this method uses a `standard length' rather than a
`standard candle' it is less dependent on physical models and of
changes in the fundamental constants, other than $c$, the speed
of light. ($c$ is needed to measure the `standard length' at the
quasar.)

\begin{figure}
\plotone{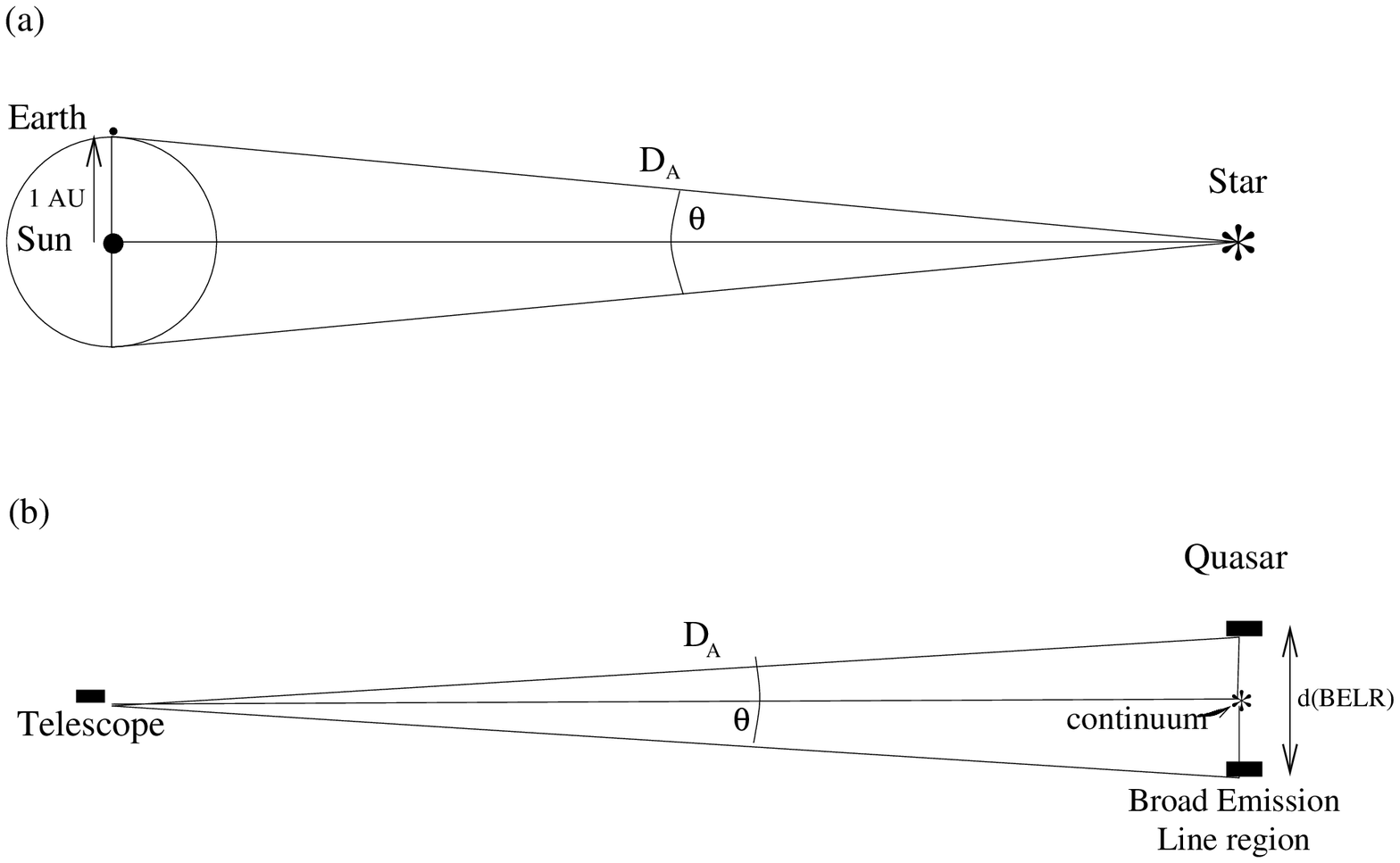}
\label{geometry}
\caption{ Measuring geometric distances to astronomical objects:
(a) parallax to a star; (b) broad emission line region distance
to a quasar.}
\end{figure}

Quasars are highly luminous objects and are readily studied in
detail up to the highest redshifts (currently $>$6, Fan et
al. 2001). There have been previous proposals to use quasars as
distance indicators (Baldwin 1977, Rudge \& Raine 1999, Collier
et al. 1999, Homan \& Wardle 2000), but these methods are, to a
greater or lesser degree, model dependent.
A defining characteristic of quasars is a series of strong broad
emission lines from permitted transitions, whose breadths
indicate Doppler velocities of 1-5\%$c$
(3000-15,000~km~s$^{-1}$).  These `broad emission lines' (BELs)
arise from gas that is photoionized by a highly luminous source
of radiation with a spectrum that is broader than any single
black body, so producing a wide range of ionization in the
BELR. Even in radio-quiet quasars, where the observed continuum
is not dominated by relativistically beamed jet emission, this
continuum is highly variable on short time scales ($\sim$1~day),
and so the source of the continuum is small, $\sim$100~AU
(1.5$\times$10$^{15}$cm) across (Peterson 1997). The response of
the BELs to continuum changes is delayed by the time light takes
to travel from the continuum source to the BEL emission region.
This light travel time gives the size of the BELR (Blandford \&
McKee 1982, Peterson 1993, 2001, Netzer \& Peterson 1997). This
technique is called `reverberation mapping'.  A typical BELR
emitting region radius in nearby active galaxies (for
CIV$\lambda$1550\AA, Netzer \& Peterson 1997) is 10~light-days
(2.5$\times$10$^{16}$~cm, Wandel, Peterson \& Malkan 1999, Kaspi
et al. 2000).

In the upper panel of figure~2 we show the angular diameter
vs. redshift, $D_A$ vs. $z$, relation for a $\Lambda$=0,
$\Omega$=1 cosmology and in a $\Lambda$=0.7, $\Omega_m$=0.3
cosmology, for a nominal size of 10~light-days.  The turn-around
in angular diameter at z$\sim$1-2 is present in both cosmologies,
(and is a feature of all $\Omega$=1 space-times)
but is notably weaker in the positive $\Lambda$ model.
The crucial value for determining $\Lambda$ is the ratio of
$D_{A}$ in the two cosmologies.  This ratio is shown in the lower
panel of figure~2. The deviation between the two $D_{A}(z)$
relations is a relatively large effect, reaching 40\% by z=2, and
then continuing to grow slowly to 55\% at z=6.

\begin{figure}
\plotone{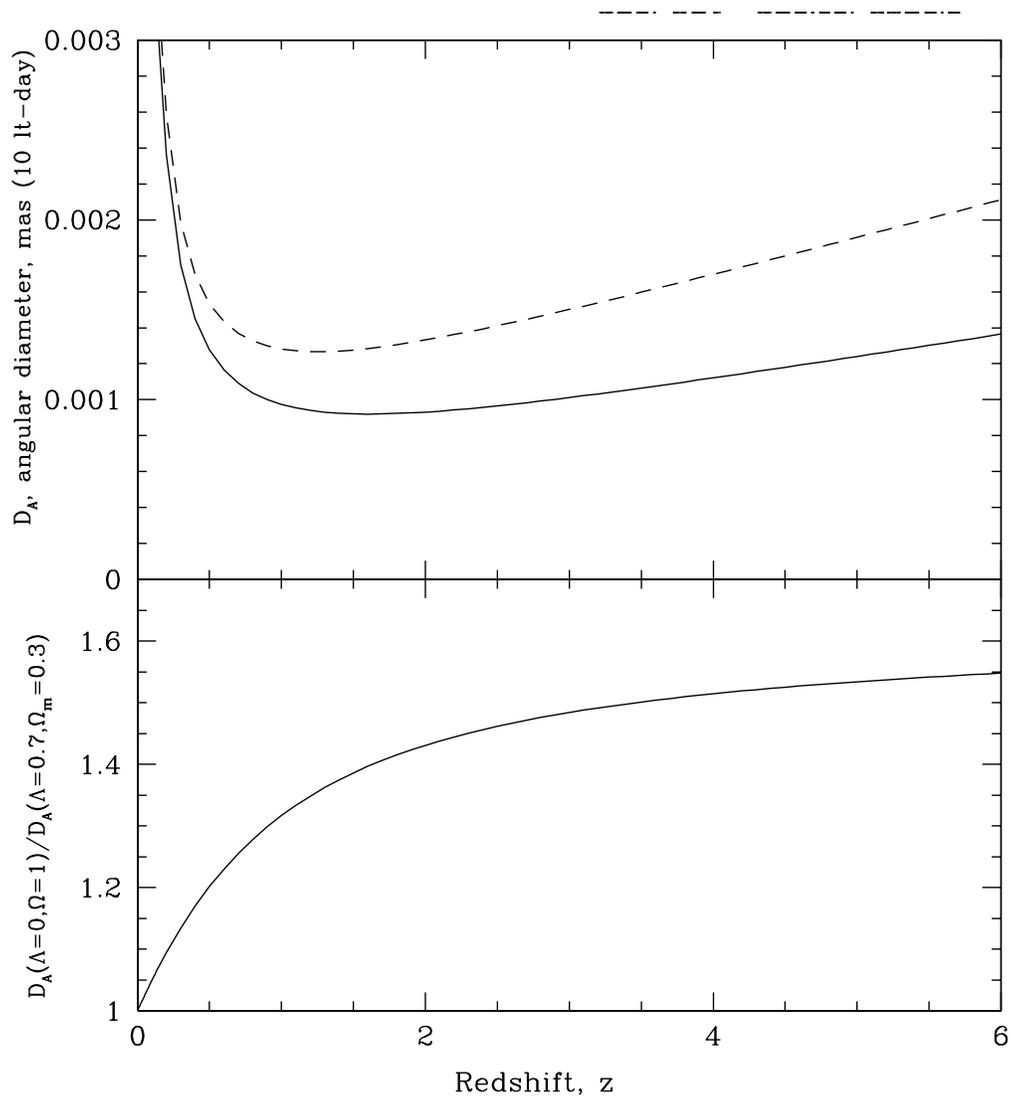}
\label{angd10ld}
\caption{
{\em top:} angular diameter of a 10~light-day long
`standard length' in milli-arcseconds for two cosmologies:
$\Omega$=1, $\Lambda$=0 (dotted line), 
$\Lambda$=0.7,  $\Omega_m$=0.3 (dot-dash line).
{\em bottom:} ratio of angular diameter in an $\Omega$=1,
$\Lambda$=0 cosmology to that in a $\Lambda$=0.7, $\Omega_m$=0.3
cosmology.}
\end{figure}

To resolve a radius of 10~lt-days at $z>$2 requires 0.001~mas
resolution (figure~2). This appears to imply a long baseline,
$d$, of 100~km at a wavelength, $\lambda$=5000\AA, using the
Rayleigh criterion ($\theta$=1.22$\lambda/d$). This is currently
a demanding requirement, though by no means an unachievable one
(Labeyrie 1999).
Fortunately quasar BELRs at z=2 are significantly larger than
those nearby. This is because the radius of the BELR, $r_{BELR}$,
increases with quasar luminosities, $L$, as $L^{0.5}$ (Wandel,
Peterson \& Malkan 1999), and $L$ evolves with redshift as
$(1+z)^3$ (Boyle et al. 1988, 2000), so $r_{BELR}\propto
(1+z)^{1.5}$. Hence at z=3 a typical quasar (i.e. one at the
break luminosity, $L^*$, in the optical luminosity function) has
a BELR 8~times larger than one at z=0, with typical {\em
diameters} at $z$=2-3 of $\sim$0.01~mas.  Although this is still
a demanding resolution, quasar evolution gives an order of
magnitude reduction in the baseline of the interferometers
required, from 5~km to 1~km at 5000\AA.

At low redshifts ($z\sim$0.2) the BELR angular diameters are some
ten times larger (0.05-0.15~mas, figure~3).  Nearby active
galaxies would not give a measurement of $\Lambda$, but would
determine $H_0$, independent of all other `distance
ladder' methods (e.g. Freedman et al. 2001).  Diameters of
$\sim$0.1~mas can be measured using a $\sim$200~meter baseline in
the UV (at CIV~1549\AA$\times$1.2), or $\sim$500~meters in the
optical (at H$\beta$4861\AA$\times$1.2).

\begin{figure}
\plotone{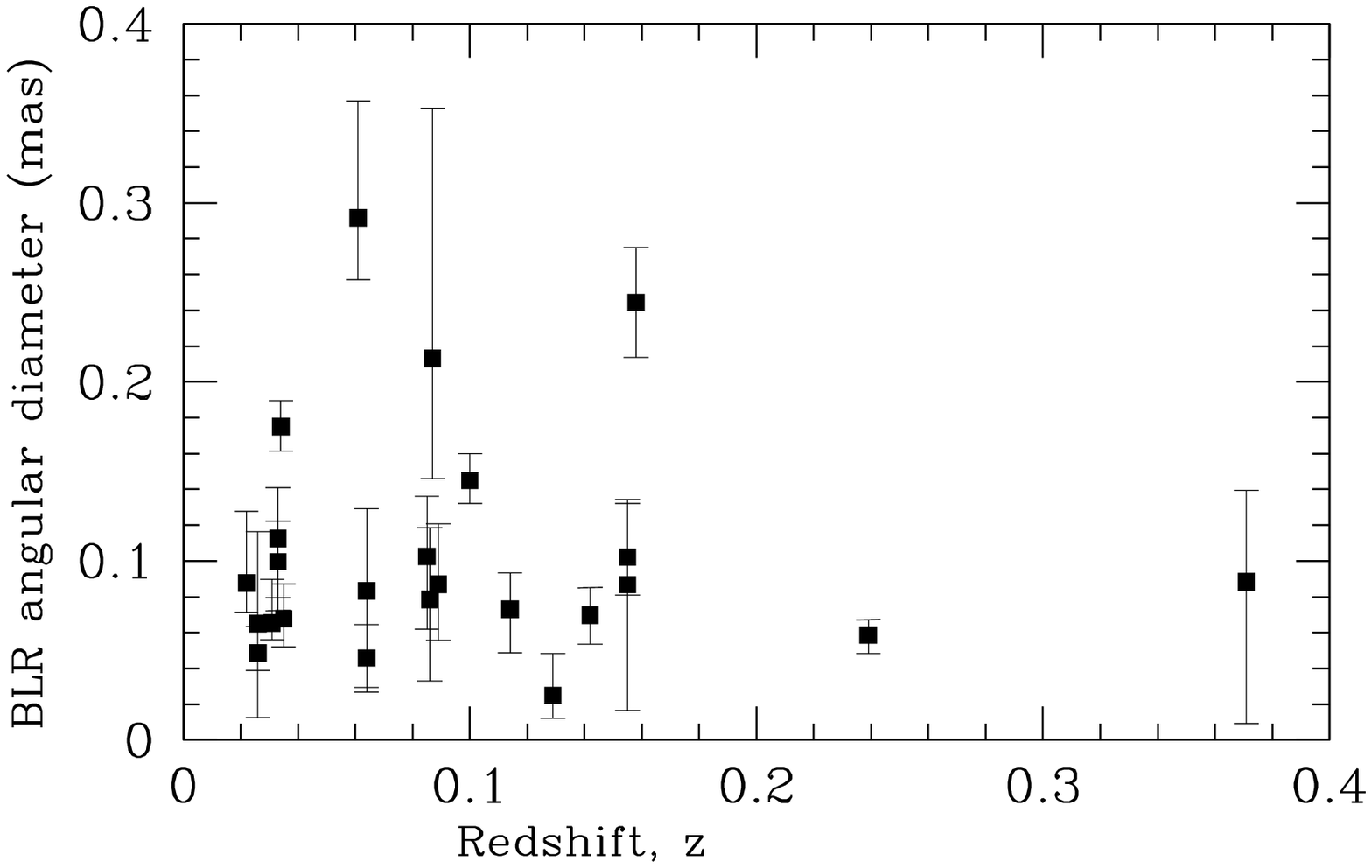}
\label{nearby}
\caption{Angular diameters for the H$\alpha$ and H$\beta$ BELRs
of nearby active galaxies, assuming
H$_0$=65~km~s$^{-1}$Mpc$^{-1}$ (Peterson et al. 1998, Kaspi et
al. 2000). (The values of $\Omega$ and $\Lambda$ are unimportant
at these low redshifts.)}
\end{figure}

The choice of strong BELs to measure is reasonably limited
(Osterbrock 1989): Ly$\alpha$ (at 1215\AA\ in the rest frame),
MgII (2800\AA), CIV (1549\AA), H$\beta$ (4861\AA), and H$\alpha$
(6562\AA).  The hydrogen lines and CIV vary quickly, while MgII
varies more slowly, in response to continuum changes.  Weaker
lines are also seen [e.g. OVI (1031\AA), He II (1640\AA,
10830\AA), H~Pa-$\alpha$ (18751\AA), H~Pa-$\beta$ (12818\AA)] and
would be candidates under special constraints of redshift and
observing band. If the strong narrow X-ray Fe-K emission line in
active galaxy spectra at 6.4~keV comes from a BELR sized region
or larger, as seems increasingly likely (Chiang et al. 2000,
Elvis 2000, Takahashi et al. 2001), then X-rays too would become
attractive for measuring $\Lambda$.

Despite this limited choice of emission lines, the optimum
wavelength band of the interferometer to use is not obvious.  The
observed wavelength of any line increases as $(1+z)$, so that the
baseline of an interferometer that can resolve the BELR increases
with $z$ for a fixed physical size.  So in general shorter
wavelength emission lines are preferred. However, when these lie
short-ward of 3000\AA, in the UV or X-ray bands, they can be
observed only from space. Several long-baseline space-based
interferometers are currently in the planning stages. These
include interferometry projects under development such as the
Stellar Imager (Carpenter, K. et al 2002), with a 0.1 mas
resolution at UV wavelengths , and long term concepts such at the
Terrestrial Planet Imager with 6000~km baselines. 
1999)
In the X-ray domain baselines $\sim$1000 times shorter than in
the optical can be used: $\sim$10~m for $\sim$0.01~mas resolution
at 2~keV (i.e. the energy of the Fe-K line at $z$=2-3). A
diffraction limited version of the {\em Chandra} X-ray
observatory could directly resolve $\sim$20~mas. Building such
a telescope may be feasible (L. Vanspeybroeck, 2002 private
communication). However, to measure $H_0$ and $\Lambda$ would
require X-ray interferometry, which is under development (Cash et
al. 2000).

From the ground, interferometry is easier at longer wavelengths
because the atmosphereric patches that limit angular resolution
are larger there. The near infrared (1-2$\mu$m, the JHK bands) is
especially promising, but requires 10 times longer baselines than
in the UV.  However, several ground-based optical and near-IR
interferometers employing large aperture telescopes (e.g. 8-m) on
baselines of several hundred meters are currently operating or
being built (VLT-I, CHARA, {\em Ohana}).  The telescope sizes
needed for interferometry may be large, since interferometers
require narrow bandwidths and these narrow bands may sample only
a fraction of the emission line profile.  (e.g. VLT-I, Petrov et
al., 2000; {\em Ohana}, Perrin et al. 2000).  Finding an optimal
practical choice of BEL, redshift interval and observing
technique is a complex problem.

Using high-signal to noise visibility measurements and a model
brightness distribution, the limiting resolution of a a telescope
can significantly exceed the Rayleigh criterion.  For example
Karovska et al. (1989) measured diameters of several stars and of
SN1987A with the CTIO 4-m telescope, and obtained angular
diameters of 5-10~mas, with uncertainty of $\pm$1~mas.  I.e. as
much as 5 times smaller than the the Rayleigh diffraction limit
($\sim$30~mas at optical wavelengths). Similarly, the HST Fine
Guidance Sensors (FGS) have been used to measure the sizes of
objects to scales of 8~mas, well below the $\sim$50~mas optical
diffraction limit of HST (e.g. Hook et al. 2000).

Accurate measurements of the reverberation times of fairly high
redshift quasars are required to carry out this distance
determination project. Since the size of the broad emission line
region depends systematically on the line measured, with lower
ionization lines coming from larger regions, the lag-time and
angular size measurements must be carried out using the same
emission line.  Moreover different velocities within a line
likely come from different locations, so the same velocity slice
of the emission lines must be observed by both techniques.

Radio-quiet quasars at z$\sim$2-3 seem to be sufficiently
variable. Optical continuum variations of at least 0.5$^m$ are
seen in at least 40\% of bright ($m<18$) radio-quiet quasars
(Netzer and Sheffer 1982, Maoz et al. 1994), although Kaspi
(2002) finds little variability in another sample.  Quasars at
z$\sim$2-3 are also sufficiently bright. Most of the difference
between $\Lambda$=0 and $\Lambda$=0.7 cosmologies has been
reached by $z$=2-3 and, as these redshifts span the peak of
quasar evolution, there are many optically bright (B=17-18~mag)
objects to observe at these redshifts (Wisotzki 2000).
The time scale of BEL variations increases with redshift as
$(1+z)^{2.5}$, since not only does the BELR size increase, but
also the intrinsic variation time scale is dilated by $(1+z)$.  As
a result, even for the fastest varying lines, e.g. CIV, to
observe for 3-5 times the typical reverberation time for quasars
at z=3 requires observing programs spanning of 3-5~years. While
this lengthens the total program, the time span is not
prohibitive, and the frequency of observations is less demanding
than in the smaller, more rapidly varying, active galaxies so far
measured by reverberation mapping (Peterson 2001).

To measure $\Lambda$ well requires a final accuracy of order 10\%
(figure~2).  The error on measured BELR radii is currently quite
large, even up to factors of 2 for the moderately large redshift
quasars in Kaspi et al. (2000). However this is not intrinsic to
the method, and the smallest error derived so far is 8\%
(Peterson et al. 1998). The lag time error is a function of
observation density, spectral signal-to-noise ratio and
calibration accuracy, and of catching a quasar in a strong
outburst. For example, the 1993 HST campaign on NGC~5548 found
only weak (75\%) continuum variations, but had the campaign
lasted twice as long ($\sim$80~days) a factor 2.5 change would
have been found, leading to much improved reverberation lag
measurements. A factor 4.5 variation was recorded in an earlier,
less intensive, campaign (Clavel et al. 1991) on the same object.
Dedicated observing facilities may be required to measure
$\Lambda$ by this method.

To achieve 5\% accuracy in lag times, and so leave room for an
8\% error in measuring $\theta$, requires sampling at 0.02-0.1
the BELR light crossing time, $r_{BELR}/c$ (Peterson et al. 2002,
Edelson et al. 2001). Several lag times must be sampled to give
an unambiguous result (Horne et al. 2002), giving a nominal
minimum of order 200 observations per quasar, i.e. of order once
per week at $z$=2-3. These estimates are probably somewhat
optimistic, and a well designed program would exceed them by a
factor of a few.
Large telescopes are not needed to measure high redshift quasar
reverberation times.  At $z$=2 the 9.9 times longer lag times
helps to compensate for the 100 times decreased flux compared
with $z$=0.2. A 1.5~meter telescope is enough to achieve a
signal-to-noise of 100 in the CIV line at this sampling rate,
even allowing for the slight weakening of the high ionization
emission lines with luminosity (Baldwin 1977).  (This estimate
applies to space-based instruments where sources of noise other
than counting statistics can be rendered unimportant). Velocity
resolved spectra, to match interferometric bands, will require
larger telescopes.

Fortunately observations indicate that a single time scale does
dominate for each emission line (Netzer \& Peterson 1997).  Even
so, there is an inherent problem in relating reverberation lag
times as currently measured to the angular sizes that an
interferometer would measure. Even perfectly sampled data will
produce a broadened cross-correlation function (CCF), and it is
not obvious which delay time, e.g. the CCF peak, or the CCF
centroid, corresponds to the angular size from an image.
Moreover, reverberation mapping measures a `responsivity
weighted' size of the light delay surface, while interferometry
measures an `emission weighted' size. That these are different is
shown by the fact that the amplitude of variation of the emission
lines is much less than that of the continuum. Quite often,
factor 2 changes in the UV continuum produce only 20\% changes in
the emission lines (e.g. Peterson 2001, figure~31). The high
ionization and high order lines vary most strongly, but they also
tend to be weak (Ulrich et al., 1991). Not only does this weak
response require larger telescopes, but also the gas being
studied is not the same in the two types of measurement.

We suggest that this difficulty can be removed by observing the
target quasar with an interferometer in both low and high
continuum states, with appropriate lags included, so that the
difference map will be a measure only of the `responsive
fraction' of the line, i.e.  that part which responds to
continuum changes. The two methods would then be measuring the
same thing. This approach would require monitoring a number of
quasars for continuum changes, and then using them to trigger
`target of opportunity' interferometer measurements.

A more ambitious approach would be to monitor quasars with a true
imaging interferometer. We could then watch continuum changes
propagating outward and creating emission line responses as a
function of wavelength.  Since the geometry of quasar BELRs is
not known this `imaging reverberation mapping' method would go
a long way to removing the ambiguities in the method.
Possibilities for the BELR geometry range from a simple orderly
wind (Elvis 2000) to a maximally chaotic distribution of clouds
(Baldwin et al. 1995). The BELR has 6 dimensions of structure: 3
of position plus 3 of velocity.  Imaging of the BELR
reverberations in several velocity bands, while the continuum
varies would provide 4 observed dimensions (RA, dec, velocity and
lag time) directly.  If there is an axis of symmetry, then the
BELR will be under-determined only by 1 dimension. The angular
diameter distances derived would then have minimal room for
uncertainty.  Currently reverberation mapping measures only 1+
dimensions (lag time, and minimal velocity information), so 4
dimensions would be a great advance.

However, even this method remains 1 dimension short of
being fully determined.  Long term (years) variations of the BELR
may provide the missing 6th dimension.  BEL profile changes occur
in some objects (e.g. Akn~120, Kollatschny et al. 1981).
Moreover, the size of the BELR for particular emission lines
seems to respond to changes in the central continuum flux. Lag
times in NGC~5548 seem to be longer when the flux is high than
when the flux is low (Peterson et al. 1999, Peterson et
al. 2002).  As a result, the size and lag measurements are best
made at the same time. Again, the best program would be the
continuous interferometric monitoring mentioned above. A fully
solved 6-D BELR structure and kinematics would not only be a
major advance in its own right, but would also minimize the
uncertainties in the cosmological parameters derived from this
method.

Although interferometry is presently not ready to make a
measurement of $\Lambda$ using quasar parallax, or perform true
imaging of quasar BELRs, ground-based interferometry has reached
a stage in which initial determinations of diameters of the BELRs
of nearby AGNs is feasible.  If the geometry of a sample of
nearby AGNs can be explored with these long-baseline
interferometers then they will lay the foundations for a long
term program to determine cosmological parameters by this
method. In the near future sub-milliarsecond resolution will be
achieved with the long baseline (200-800~m) interferometers VLT-I
and Ohana in the near-IR, where the Paschen lines are
available at low z. In the following few years, imaging should
become possible in the optical as well (e.g. at H$\alpha$).
Imaging of nearby AGN BELRs in the lines used in reverberation
measurements will determine their spectral/spatial morphology,
and allow the technique to be refined.
This will provide a crucial base for `parallax' measurements of
high redshift quasars to determine $\Lambda$, when the next
generation of long-baseline interferometers with $<$0.01~mas
resolution become available.


\bigskip

We thank Brad Peterson and Rick Edelson for valuable comments on
reverberation mapping measurements, Jill Bechtold for the use of
her $\Lambda$ cosmology code, Jonathan McDowell for assistance
with general relativity, and Helene Sol for informing us of the
{\em Ohana} project. We thank the anonymous referee for comments
on line responsivity which led to a significant improvement in
the paper. This work was supported in part by NASA contract
NAS8-39073 (Chandra X-ray Center).

\newpage


\end{document}